# Accelerating Spiking Neural Networks Using Quantum Algorithm with High Success Probability and High Calculation Accuracy

Yanhu Chen[1], Cen Wang[2], Hongxiang Guo[1]*, Xiong Gao[1], Jian Wu[1]

1 Institute of Information Photonics and Optical Communications, Beijing University of Posts and Telecommunications, Beijing 100876, China

2 Photonic Transport Network Laboratory, KDDI Research Inc., Saitama-ken 356-8502, Japan

**Abstract :** Utilizing quantum computers to implement artificial neural networks (ANNs) will obtain the potentially significant advancements of both computing speed and model scale. In this paper, we propose a quantum algorithm to implement a quantum spiking neural network (QSNN). We give a mathematical proof to show that using our quantum algorithm, the computational complexity of the QSNN is a log-polynomial relationship with the data dimension, which is much lower than that of the classic spiking neural network (SNN) (i.e., implemented on a classic computer). We also comprehensively evaluate the performance of the QSNN including its success probability, calculation accuracy. Furthermore, we provide a method to improve the minimum success probability of the QSNN to nearly 100%. Finally, we verify the feasibility of the QSNN by solving a real-world image classification task.

## 1. Instruction

Deep learning usually based on artificial neural networks (ANNs) has obtained remarkable achievement in both research and industry fields span on pattern recognition, image classification, and decision making [1, 2]. However, with the explosive growth of training data and model complexity, ANNs that require massive computing have made classic computers be close to the ceiling of their computing power. Additionally, quantum tunneling effect of semiconductors on the microscopic scale [3] and the lower bound of energy consumption (the Landauer's principle) [4] may cause that the legacy computer efficiency cannot follow Moore's Law to continuously obtain rapidly improvements [5]. Consequently, a novel hardware is expected, and the quantum computers is thought to be a promising choice among numerous computing platforms. For the reason that quantum has the natures of superposition and entanglement, in recent years, multi-type ANNs have been implemented on quantum computers (i.e., the details are illustrated in Section 2) to reduce the computational complexity and further expand the scale of neural networks [6-17].

With the development of neuroscience [18, 19], another ANN designed according to the biological nervous system, the spiking neural networks (SNNs) were proposed [20]. Due to its nature of neuromorphic computing [21, 22], SNN is often regarded as the next generation neural architecture and can enable a higher-level intelligence such as brain-like learning [23]. Specifically, a spiking neuron in an SNN can emulate a biological neuron. Responding to a series of temporal-spatial stimuli, a potential of a spiking neuron will undulate dynamically. The spiking neuron only fires when a potential over the threshold happens. Then, the fired spiking neuron generates an output stimulus. This output stimulus can be retrieved or transmitted to other connected spiking neurons [21, 24, 25]. It can be seen that SNNs incorporate the concept of time into their operating model, so that SNNs can process the temporal-spatial data.

However, using quantum algorithm to accelerate the SNNs was always overlooked in previous studies on quantum ANNs. Inspired by the related works, in this paper, in order to further speed up the classic SNN, we propose a quantum algorithm so that SNN can be executed on a quantum computer. Specifically, to accelerate implementation of SNN, we propose a quantum algorithm for a core process that cost maximum computing power. This process is to find all the moments of the spiking neurons that their potentials cross the threshold. Through mathematical derivation, we verify that the core process can be transformed into the operation of vector inner products, and we design the quantum circuits to efficiently solve it.

Our contributions are described as follows: (1) To our best of the knowledge, we are the first to propose a QSNN, using quantum algorithm to accelerate the classic SNN, which can adapt learning requirements with arbitrary data dimension and data accuracy. (2) We give the mathematical proof on the originally minimum success probability and calculation accuracy of SNN implemented on a quantum computer which are the key performance indicators of a quantum algorithm, and then provide an approach to further improve the minimum success probability of QSNN. (3) Based on the analysis of quantum gate complexity combined with the minimum success probability, we verify that the computational complexity of QSNN is log-polynomial, which is much lower than the linear computational complexity of the classic SNN. (4) In order to validate the feasibility of the quantum algorithm, we also apply the QSNN to solve a real-world image classification task.

## 2. Related work

In this section, we review various kinds of quantum neural networks and their applications.

Dating back to 1995, KaK et.al first proposed the concept of quantum neural computation. They suggested that it is a novel research field to combine neural computing and quantum computing to form a new paradigm [6]. In 2000, Narayanan et.al proposed quantum artificial neural network (QANN) architectures and compared the performance of quantum/classic neural networks [7]. They concluded that the QANN has overwhelming advantages in both computing speed and model scale. Since then, deep learning models based on quantum computing are actively and widely studied.

Mainstream researches of QANNs are to implement feed-forward neural networks by designing quantum circuits. Kouda et.al proposed to use the quantum phase for the implementation of the qubit neural network and discussed its learning efficiency [8]. Tacchino et.al proposed a quantum perceptron [12] and further constructed a multilayer quantum neural network based on the quantum perceptron [13]. Killoran et.al proposed a continuous-variable QANN [14]. Beer et.al proposed an approach for training deep QANNs [17]. In addition, as the modern neural networks vary in architectures, some researchers also explored quantum versions of the special architectures. For example, Cong et.al proposed a quantum convolutional neural network (QCNN). QCNN can be used not only in the classic pattern recognition, but also in devising a novel quantum error correction scheme [15]. Rebentrost et.al proposed a quantum Hopfield neural network (QHNN), QHNN can be used to store quantum data and even recognize virus RNA sequences [26]. Pierre-Luc et.al proposed a quantum generative adversarial network (QGAN) [11]. Then, Zoufal et.al used QGAN for learning and loading quantum states of random amplitude in polynomial complexity with the number of qubits [16]. Romero et.al proposed a quantum autoencoder to efficiently compress the quantum data [10]. In order to build a more generalized QANN model, Benedetti et. al creatively introduced the gradient descent method to update the parameters of the quantum circuits [27].

## 3. Review of SNN Architecture

SNN has a significant difference in information processing within a neuron compared to the traditional ANNs. In ANNs, the inputs and outputs of a neuron are real-values. While in SNNs, a spiking neuron will generate stimuli sequences according to the input stimuli sequences [22]. A stimulus is essentially a binary event, so that a stimuli sequence is a 0/1 sequence. A basic architecture of a SNN is a two-layer architecture that consists of $N$ synapses and a spiking neuron, which is shown in Fig 1. Given an observation period $T$, the synapses receive the input stimuli sequences and convert the stimuli into the spikes. The spiking neuron is connected to the synapses, and each connection has a tunable weight. During $T$, the potential of the spiking neuron is the weighted sum of the spikes on the synapses and can be described as in equation (1).

$$V(t) = \sum_k w_k \sum_{t_i} K(t - t_i) + V_{rest} \quad (1)$$

where $w_k$ denotes the weight of the $k$-th synapse; $t_i$ denotes the moment that the $i$-th spike from the $k$-th synapse is received by the spiking neuron; $V_{rest}$ denotes the rest potential of the spiking neuron, defaulted $V_{rest} = 0$; function $K(t - t_i)$ emulates the spike shape to respond a received stimulus, which is described in equation (2):

$$K(t - t_i) = V_0 \left( e^{\frac{-(t-t_i)}{\tau}} - e^{\frac{-(t-t_i)}{\tau_s}} \right) \quad (2)$$

where $\tau$ and $\tau_s$ are hyper-parameters and denote the decay time; $V_0$ is a normalized parameter, and $V_0 = \frac{\tau_s \tau}{\tau - \tau_s}$ [24]. Finally, only when the potential $V(t^*)$ is over the threshold, the spiking neuron will produce an output stimulus at $t^*$. In Fig. 2, we give an example to intuitively display the potential variation of the spiking neuron.

It can be seen that in $T$, the input stimuli sequences from multiple synapses can be reproduced to a new output stimuli sequence. The output stimuli sequence can be read out directly or can be regarded as the input stimuli of a synapse in a next layer. Following the same manner, a more general SNN that has multiple layers can be easily extended by the two-layer SNN.

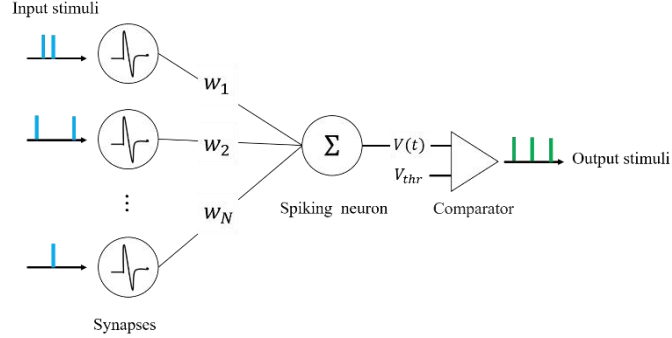

Fig. 1. A simple two-layer SNN architecture.

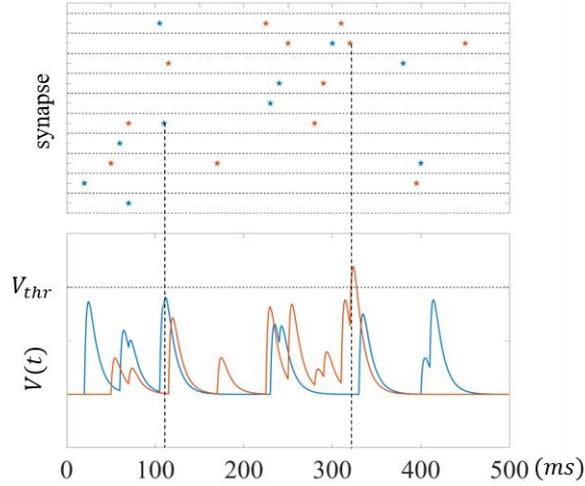

Fig 2. The potential variations of a positive case (i.e., with threshold crossing) and a negative case (i.e., without threshold crossing), which are drawn by red and blue color, respectively, in a spiking neuron during the observation period. The localization of a star denotes a moment that the corresponding synapse receives a stimulus.

## 4. Constructing of QSNN

According to the general SNN model, we find that the part with highest computational complexity of SNN is to locate the moment that the spiking neuron potential crosses the threshold. Through mathematical derivation, we verify that the localizations of the moments that the spiking neuron potential crosses the threshold can be transformed into the calculation of the unsigned inner products of two vectors. Therefore, our QSNN aims to accelerate this process to obtain efficiency optimizations. Then, we further design the scalable quantum circuits of QSNN, which can adapt learning requirements with arbitrary data dimension and data accuracy. We also evaluate the success probability and calculation accuracy of QSNN. We then provide an approach to further improve the minimum success probability. Additionally, based on the gate complexity of the quantum circuits and the minimum success probability, we illustrate that QSNN can obtain an exponential acceleration compared to the classic SNN.

## 4.1 Localization of Threshold Crossing

In order to obtain the output stimuli, we must know whether $V(t)$ crosses $V_{thr}$. The most intuitive way is using the derivative of $V(t)$ with respect to $t$ to obtain all the local maximum potentials $V(t_{max})$ of the spiking neuron and their corresponding moments $t_{max}$, then compare $V(t_{max})$ with $V_{thr}$.

Assuming that the total number of stimulus in an input stimuli sequence is $J_a$, the set of the arrived moments of the stimuli is $D = \{t_1, t_2, \cdots t_J \cdots t_{J_a}\}$ (any $t_J \leq T$) and the number of elements in $D$ is $\|D\|$. Because an arrived stimulus at $t_J$ does not affect the spiking neuron potential before $t_J$, we cannot directly seek the derivative of $V(t)$ in equation (1). However, we can seek a derivative of $V_J(t)$ in equation (3), which is a correct operation for the first $J$ stimuli. Therefore, we actually need to compute $\|D\|$ derivative operations.

$$V_J(t) = \sum_{j=1}^{J} w_j K(t - t_j) \quad (3)$$

where $w_j$ and $t_j$ denote the weight and the moment of the $j$-th stimulus; Given the function $K(\cdot)$ in equation (2), the derivative of $V_J(t)$ with respect to $t$ is shown as equation (4):

$$\frac{dV_J(t)}{dt} = V_0 \sum_{j=1}^{J} w_j \left[ \frac{1}{\tau_s} e^{-\frac{t-t_j}{\tau_s}} - \frac{1}{\tau} e^{-\frac{t-t_j}{\tau}} \right] \quad (4)$$

Thus, under the $J$ input stimuli, the moment $t_J$ when the potential on the spiking neuron reaches the local maximum is:

$$t_J = V_0 \left[ \ln \frac{\tau}{\tau_s} + \ln \left( \sum_{j=1}^{J} w_j e^{\frac{t_j}{\tau_s}} \right) - \ln \left( \sum_{j=1}^{J} w_j e^{\frac{t_j}{\tau}} \right) \right] \quad (5)$$

Now, we can find that the summation operation in equation (5) could be rewritten using vectors, as shown in equation (6); thus, to find can be converted to an operation of calculating the inner product of two vectors, as shown.

$$t_J = V_0 \left[ \ln \frac{\tau}{\tau_s} + \ln \left( \vec{w}^T \vec{t}_{\tau_s} \right) - \ln \left( \vec{w}^T \vec{t}_{\tau} \right) \right] \quad (6)$$

where $\vec{w} = [w_1, w_2, \ldots, w_J]^T$, $\vec{t}_{\tau_s} = \left[ e^{\frac{t_1}{\tau_s}}, e^{\frac{t_2}{\tau_s}}, \ldots, e^{\frac{t_J}{\tau_s}} \right]$ and $\vec{t}_{\tau} = \left[ e^{\frac{t_1}{\tau}}, e^{\frac{t_2}{\tau}}, \ldots, e^{\frac{t_J}{\tau}} \right]$. Two constraints, $\vec{w}^T \vec{t}_{\tau_s} > 0$ and $\vec{w}^T \vec{t}_{\tau} > 0$, should be satisfied due to the $\ln(\cdot)$ function. Thus, the inner product of vector $\vec{w}$ and vector $\vec{t}_{\tau_s}$ or $\vec{t}_{\tau}$ must be unsigned.

From equation (6), we can see that to obtain $t_J$ requires higher massive computing powers as the number of input stimuli increasing. As shown in equation (6), to calculate $t_J$, we need to obtain the unsigned inner product of $\vec{w}$ and $\vec{t}_{\tau_s}$, $\vec{w}$ and $\vec{t}_{\tau}$. To reduce computational complexity of this part which costs highest computing powers in the classic SNN, and then to obtain further accelerations accordingly, we propose a **q**uantum algorithm to compute the **u**nsigned **i**nner **p**roduct of vectors (QUIP) and then use this algorithm to construct QSNN.

## 4.2 QUIP

The origin quantum approach to compute the unsigned inner product is swap-test [28]. The swap-test consists of three parts, an ancillary qubit and two quantum registers, as shown in Fig 3. $U_t$ and $U_w$ can transform $|0\rangle^{\otimes \lceil \log_2 J \rceil}$ to the two vectors $|w\rangle$, $|t\rangle$, and the inner product of these two vectors are needed to be calculated:

$$U_w |0\rangle_w = |w\rangle \quad (7)$$
$$U_t |0\rangle_t = |t\rangle \quad (8)$$

where $|w\rangle = \left(\sum_{j=0}^{J-1} w_j^2\right)^{-\frac{1}{2}} \vec{w}$, $|t\rangle = \left(\sum_{j=0}^{J-1} e^{2\frac{t_j}{\tau_s}}\right)^{-\frac{1}{2}} \vec{t}_{\tau_s}$ or $|t\rangle = \left(\sum_{j=0}^{J-1} e^{2\frac{t_j}{\tau}}\right)^{-\frac{1}{2}} \vec{t}_\tau$. In the swap-test, an ancillary qubit is introduced. We will obtain the quantum state $|\Psi\rangle$, after performing the swap-test.

$$|\Psi\rangle = \frac{1}{2}|0\rangle_{anc}(|w\rangle|t\rangle + |t\rangle|w\rangle) + \frac{1}{2}|1\rangle_{anc}(|w\rangle|t\rangle - |t\rangle|w\rangle) \quad (9)$$

where $|0\rangle_{anc}$ and $|1\rangle_{anc}$ are the two states of the ancillary qubit. The state probabilities of the ancillary qubit are $P(|0\rangle) = \frac{1}{2} + \frac{1}{2}|\langle w|t\rangle|^2$, $P(|1\rangle) = \frac{1}{2} - \frac{1}{2}|\langle w|t\rangle|^2$. We find that the state probabilities of the ancillary qubit contain our expected inner product $\langle w|t\rangle$.

Because the swap-test uses $U_w$, $U_t$ to prepare the initial states of $|w\rangle$, $|t\rangle$ and $\lceil\log_2 J\rceil$ controlled-swap gates to compute the inner product. Inspired by the parameterized quantum circuits and generative adversarial network model [16], the complexity of preparing the initial quantum is polynomial relationship with the number of qubits. Therefore, the complexity of the swap-test is $O(\text{poly}\lceil\log_2 J\rceil)$. In classic computers, computing the inner product requires $J$ multiplications, so the complexity is $O(J)$. However, the measurement operator will cause the collapse of ancillary qubit, and only 0 or 1 is the measurement result. To obtain the probability of each ancillary qubit state, we must repeatedly execute the swap-test and count the measurement result. In fact, if the required calculation accuracy of the state probability is an $\gamma$-digit binary floating-point number, the swap-test will be run and measured $4^\gamma$ times (i.e., see Appendix A for the mathematical proof). These repetitions undoubtedly cancel the advantages of a quantum algorithm for computing inner products.

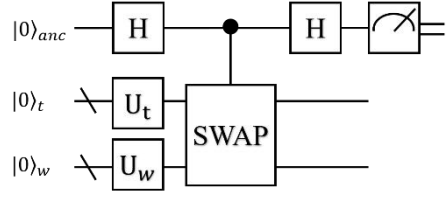

Fig 3. The quantum swap-test circuit. The top qubit is an ancillary qubit, the next two are two quantum registers.

Considering these, to avoid the repeated executions, we propose QUIP in which the core idea is to combine the quantum amplitude estimation (QAE) algorithm [29, 30] and the swap-test algorithm. Fig 4 then shows the QUIP circuits containing a control register and a target register with $m$ and $1 + 2\lceil\log_2 J\rceil$ qubits, respectively.

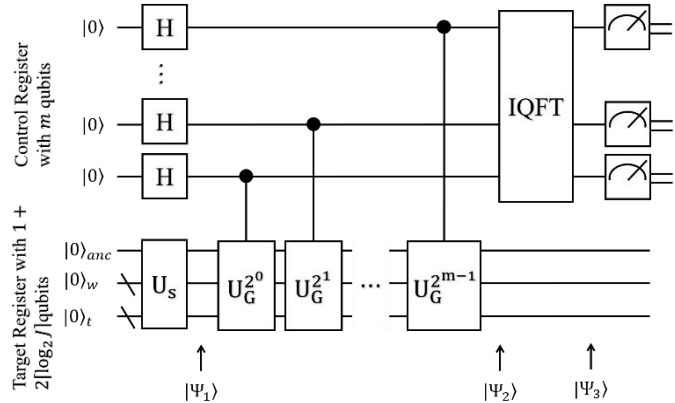

Fig. 4. The QUIP circuits for calculating and estimating $\langle w|t\rangle$. $U_s$ denotes the swap-test operator; $U_G$ denotes the generalized Grover operator; IQFT denotes the inverse quantum Fourier transformer.

The specific steps of QUIP are as follows.
(1) Because $P(|0\rangle) + P(|1\rangle) = 1$, we here use a trick to parameterize the inner products using to $\theta$, in this way we can obtain the specific expressions of $\langle w|t\rangle$. We define $P(|0\rangle) = \cos^2\theta$ and $P(|1\rangle) = \sin^2\theta$. Thus, after performing $U_s$, the quantum state of the target register is $|\Psi\rangle$ which equals to $[\cos\theta, \sin\theta]^T$. Further, in equation (10), we construct the mapping relationship between $\theta$ and $\langle w|t\rangle$:

$$\langle w|t\rangle = \sqrt{\cos 2\theta} \quad (10)$$

where $\theta \in \left[0, \frac{\pi}{4}\right]$. This is because $P(|0\rangle) = \frac{1}{2} + \frac{1}{2}|\langle w|t\rangle|^2$ and $P(|1\rangle) = \frac{1}{2} - \frac{1}{2}|\langle w|t\rangle|^2$ are probabilities between 0 and 1.

(2) As shown in Fig 4, $|\Psi_1\rangle$ consists of an equal superposition state and $|\Psi\rangle$.

$$|\Psi_1\rangle = \frac{1}{\sqrt{2^m}}\left(\sum_{k=0}^{2^m-1}|k\rangle\right)|\Psi\rangle \quad (11)$$

(3) $|\Psi_2\rangle$ is obtained by applying several controlled $U_G$ operators on $|\Psi_1\rangle$, as shown in equation (12):

$$|\Psi_2\rangle = \frac{1}{\sqrt{2^m}}\left(\sum_{k=0}^{2^m-1}|k\rangle U_G^k|\Psi\rangle\right) \quad (12)$$

where $U_G$ denotes a generalized Grover operator [30, 31]. $U_G$ can be further divided into four sub-operators, the oracle operator $O$, the conditional phase shift operator $I_0$, the swap-test operator $U_s$ and its inverse operator $U_s^{-1}$. A specific $U_G$ operator is depicted in Fig 5. $U_G$ operator can be written as:

$$U_G = -U_s I_0 U_s^{-1} O = \begin{bmatrix} \cos 2\theta & \sin 2\theta \\ -\sin 2\theta & \cos 2\theta \end{bmatrix} \quad (13)$$

(4) Furthermore, the quantum phase kick-back has to be mentioned [32]. That is, if the input quantum state of the target register is one of the eigenvectors of the controlled operator, the target register will not change and the control register will receive a phase that is equal to an eigenvalue of the corresponding eigenvector. The eigenvectors of $U_G$ are written as $|\mu_1\rangle = [1, -i]^T$, $|\mu_2\rangle = [1, i]^T$, so that $|\Psi\rangle = e^{i\theta}|\mu_1\rangle + e^{-i\theta}|\mu_2\rangle$. Therefore, $|\Psi_2\rangle$ is rewritten as:

$$|\Psi_2\rangle = \frac{e^{i\theta}}{\sqrt{2^m}}\sum_{k=0}^{2^m-1}e^{i2k\theta}|k\rangle|\mu_1\rangle - \frac{e^{-i\theta}}{\sqrt{2^m}}\sum_{k=0}^{2^m-1}e^{-i2k\theta}|k\rangle|\mu_2\rangle \quad (14)$$

(5) Applying IQFT to $|\Psi_2\rangle$, we can obtain $|\Psi_3\rangle$:

$$|\Psi_3\rangle = e^{i\theta}\left|\frac{2^m\theta}{\pi}\right\rangle|\mu_1\rangle + e^{-i\theta}\left|2^m\left(1-\frac{\theta}{\pi}\right) \bmod 2^m\right\rangle|\mu_2\rangle \quad (15)$$

Ideally, when $2^m\theta/\pi$ is an integer, $\left|\frac{2^m\theta}{\pi}\right\rangle$ is a column vector, and its $2^m\theta/\pi$-th element is 1, others are 0. If $2^m\theta/\pi$ is an integer, then $2^m\left(1-\frac{\theta}{\pi}\right) \bmod 2^m$ is also an integer. If $2^m\theta/\pi$ is not an integer, equation (15) will no longer be correct. We will discuss a general case in Section 4.3.

(6) Finally, we measure the control register. The measurement result may be one of the two values with equal probability, and a $r$ is a binary value with $m$ bits.

$$r = 2^m\frac{\theta}{\pi} \quad \text{or} \quad r = 2^m\left(1-\frac{\theta}{\pi}\right) \bmod 2^m \quad (16)$$

Because $\theta \in \left[0, \frac{\pi}{4}\right]$, except for $\theta = 0$, $2^m\theta/\pi$ and $2^m\left(1-\frac{\theta}{\pi}\right) \bmod 2^m$ have no overlapping values, which means that there is no ambiguity in one measurement. Thus, we can construct the mapping relationship between $r$ and $\langle w|t\rangle$, as described in equation (17):

$$\langle w|t\rangle = \begin{cases} \sqrt{\cos\frac{r\pi}{2^{m-1}}} & 0 \leq r \leq 2^{m-2} \\ \sqrt{\cos\left(2\pi - \frac{\pi r}{2^{m-1}}\right)} & 3 \times 2^{m-2} \leq r \leq 2^m - 1 \end{cases} \quad (17)$$

In summary, in QUIP, according to the measurement of the quantum control register, we can obtain the result $r$, a binary integer with length $m$. Based on the equation (17), we can precisely calculate the $\langle w|t\rangle$, when $2^m\theta/\pi$ can be perfectly expressed as a binary integer with length $m$. In this way, we can efficiently obtain $\langle w|t\rangle$ through QUIP, while avoid repeated executions of swap-test and statistical measurement results, thereby accelerating the SNN model.

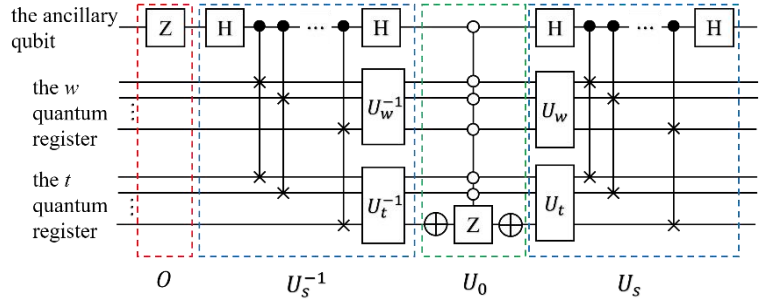

Fig. 5. The quantum circuits of the generalized Grover operator.

## 4.3 Success Probability and Calculation Accuracy

Although we have given the mapping relationship between $r$ and $\theta$, it is not perfect, since we require the measurement result $r$ must be an integer. However, since $\theta$ changes consistently in $\left[0, \frac{\pi}{4}\right]$, and $2^m \theta / \pi$ before measurement is not always an integer. The error between the actual $r$ and the measurement $\tilde{r}$ will propagate to the $\langle w | t \rangle$. It is necessary to analyze the error between the actual $r$ and the measurement $\tilde{r}$. Since the analyzations based on $2^m \left(1 - \frac{\theta}{\pi}\right) \mod 2^m$ and $2^m \theta / \pi$ are similar, we only do that for $r = 2^m \theta / \pi$.

For sake of the generality, we assume that an estimated phase $\tilde{\theta}$ and then get $\tilde{r} = 2^m \tilde{\theta} / \pi$, where $\tilde{r}$ is a binary integer with $m$ bits. The actual value $r$ is certainly not a binary integer with length $m$, due to the continuous variations of phase $\theta$. We use a remainder $\Delta r$ to represent the error term between $\tilde{r}$ and actual $r$, as shown in equation (18):

$$r/2^m = \tilde{r}/2^m + \Delta r = (0.r_1 r_2 ... r_m)_2 + \Delta r \quad (18)$$

where $r_j$ is a binary value and $r_j \in \{0,1\}$; the range of $\Delta r$ is $\left[-2^{-(m+1)}, 2^{-(m+1)}\right)$ (i.e., $\tilde{r} = \text{round}(r)$). Thus, we use the variable $r$ to replace $\theta$ in the equation (14) and get equation (19).

$$|\Psi_2\rangle = \frac{e^{i\pi r/2^m}}{\sqrt{2^m}} \sum_{k=0}^{2^m-1} e^{i2\pi rk/2^m} |k\rangle |\mu_1\rangle + \frac{e^{-i\pi r/2^m}}{\sqrt{2^m}} \sum_{k=0}^{2^m-1} e^{-i2\pi rk/2^m} |k\rangle |\mu_2\rangle \quad (19)$$

For simplification, because the two parts in $|\Psi_2\rangle$ are similar and the global phase does not affect the measurement, we only choose the left part of $|\Psi_2\rangle$ and ignore the global phase. Then, the quantum state $|S\rangle$ is described as equation (20).

$$|S\rangle = \frac{1}{\sqrt{2^m}} \sum_{k=0}^{2^m-1} e^{i2\pi rk/2^m} |k\rangle \quad (20)$$

The $2^m$ dimensional IQFT operator, $IQFT_m$, can be described as equation (21):

$$IQFT_m = \frac{1}{2^{m/2}} \begin{bmatrix} e^{-i2\pi(0\times 0)/2^m} & e^{-i2\pi(0\times 1)/2^m} & \cdots & e^{-i2\pi(0\times(2^m-1))/2^m} \\ e^{-i2\pi(1\times 0)/2^m} & e^{-i2\pi(1\times 1)/2^m} & \cdots & e^{-i2\pi(1\times(2^m-1))/2^m} \\ \vdots & \vdots & \ddots & \vdots \\ e^{-i2\pi((2^m-1)\times 0)/2^m} & e^{-i2\pi((2^m-1)\times 1)/2^m} & \cdots & e^{-i2\pi((2^m-1)\times(2^m-1))/2^m} \end{bmatrix} \quad (21)$$

Now, applying $IQFT_m$ to $|S\rangle$, we can obtain the quantum state $|r\rangle$:

$$|r\rangle = IQFT_m |S\rangle = \frac{1}{2^m} \left[ \sum_{k=0}^{2^m-1} e^{i2\pi(rk-0\times k)/2^m}, \sum_{k=0}^{2^m-1} e^{i2\pi(rk-1\times k)/2^m}, \cdots, \sum_{k=0}^{2^m-1} e^{i2\pi(rk-(2^m-1)\times k)/2^m} \right]^T \quad (22)$$

From the geometric perspective, all column vectors of $IQFT_m$ can form a group of orthogonal bases, and $|r\rangle$ is the projections of $|S\rangle$ on all orthogonal bases of $IQFT_m$. According to equation (22), in the case of $\Delta r = 0$, the vector $|S\rangle$ is parallel to the $r$-th orthogonal basis and $|S\rangle$ is orthogonal to other bases. This ideal case has been displayed in equation (15). Thus, the $r$-th element of $|r\rangle$ is 1 and others are 0 which means the measurement result is fixed. In another case, if $\{\Delta r \neq 0 \text{ and } \Delta r \in \left[-2^{-(m+1)}, 2^{-(m+1)}\right)\}$, $|S\rangle$ is not orthogonal or parallel to any orthogonal basis, but $|S\rangle$ has projections on all orthogonal bases. The square of projection length on the $h$-th orthogonal basis is the probability of the measurement $h$. Therefore, the measurement result of $|r\rangle$ may be any integer from $[0, 2^m - 1]$. Let the probability of the measurement $h$ be $p_h$, which is described in equation (23).

$$p_h = |\langle h|S\rangle|^2 = \frac{1}{2^{2m}} \left| \left( \sum_{k=0}^{2^m-1} e^{-i2\pi(h\times k)/2^m} \langle k| \right) \left( \sum_{k=0}^{2^m-1} e^{i2\pi(r\times k)/2^m} |k\rangle \right) \right|^2 = \frac{1}{2^{2m}} \left| \sum_{k=0}^{2^m-1} e^{-i2\pi(h-r)k/2^m} \right|^2 \quad (23)$$

It is obvious to see that the summation in right part of equation (26) is a geometric series. According to the summation formula of the geometric series, we can further simplify $p_h$ in equation (24).

$$p_h = \frac{1}{2^{2m}} \left| \frac{1 - e^{-i2\pi(h-r)}}{1 - e^{-i2\pi(h-r)/2^m}} \right|^2 \quad (24)$$

Based on $p_h$, we can define the success probability and the calculation accuracy for QUIP:

**Definition 1 (the success probability)**: The success probability $p_{\tilde{r}}$ is defined as the probability which the measurement result of quantum control register is $\tilde{r}$, as shown in equation (25).

$$p_{\tilde{r}} = \frac{1}{2^{2m}} \left| \frac{1 - e^{-i2\pi\Delta r \times 2^m}}{1 - e^{-i2\pi\Delta r}} \right|^2 \quad (25)$$

**Definition 2 (the calculation accuracy)**: The calculation accuracy $\varepsilon$ is defined as an absolute value of using the actual local maximum potential $\langle w|t\rangle$ minus the estimated local maximum potential $\widetilde{\langle w|t\rangle}$, when the measurement result is $\tilde{r}$, as shown in equation (26).

$$\varepsilon = \left| \langle w|t\rangle - \widetilde{\langle w|t\rangle} \right| \quad (26)$$

We first discuss $p_{\tilde{r}}$. Because the formats of the numerator and denominator in equation (25) are similar, we take the denominator as a case to show the trick for simplifying $p_{\tilde{r}}$. We treat the denominator as a vector difference on the complex number field which is highlighted in red in Fig 6(a). Besides, because the two vectors, 1 and $e^{-i2\pi\Delta r}$, have the same length, the two vectors and their difference together form an isosceles triangle. The value of the denominator equals to the vector difference, and can be simply calculated as shown in Fig 6(b). In this way, we can give the simplified measurement probability $p_{\tilde{r}}$ when the $\Delta r \neq 0$ and $\Delta r \in [-2^{-(m+1)}, 2^{-(m+1)})$, as described in equation (27).

$$p_{\tilde{r}} = \frac{1}{2^{2m}} \frac{\sin^2(2^m \pi \Delta r)}{\sin^2(\pi \Delta r)} \quad (27)$$

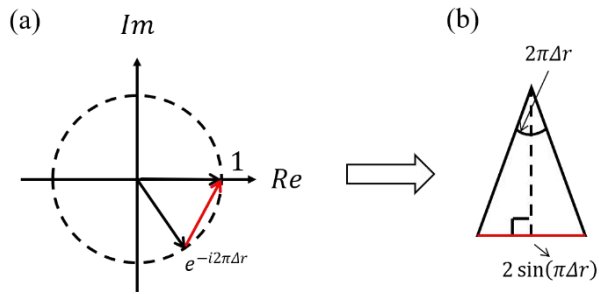

Fig 6. The diagram of symbolic-graphic combination for explaining the simplified process of $p_{\tilde{r}}$.

Then, we try to find the minimum success probability, which is an indicator of the feasibility of QSNN. We can get the minimum success probability using the derivative of $p_{\tilde{r}}$.

$$\frac{dp_{\tilde{r}}}{d\Delta r} = \frac{\pi \sin^2(2^{m+1} \pi \Delta r)}{2^m \sin^2(\pi \Delta r)} - \frac{2\pi \sin^2(2^m \pi \Delta r) \cos(\pi \Delta r)}{2^{2m} \sin^3(\pi \Delta r)} \quad (28)$$

Considering the range of $\Delta r$, we can obtain the minimum $p_{\tilde{r}}$ in equation (28).

$$\min p_{\tilde{r}} = \frac{1}{2^{2m}} \frac{\sin^2(2^m \pi \Delta r)}{\sin^2(\pi \Delta r)} \approx \frac{\sin^2(2^m \pi \Delta r)}{2^{2m} (\pi \Delta r)^2} = \frac{4}{\pi^2} \quad (29)$$

Next, we discuss $\varepsilon$. The error $\varepsilon$ is related to $\tilde{r}$ and thus has a relationship with $\Delta r$, when $m$ is deterministic. Similarly, by computing $\frac{d\varepsilon}{d\Delta r}$, we can obtain the maximum error, when $\Delta r \to 2^{-(m+1)}$ or $\Delta r \to -2^{-(m+1)}$.

$$\max \varepsilon = |\langle w|t\rangle - \widetilde{\langle w|t\rangle}| = \left|\sqrt{\cos\left(\frac{\pi\tilde{r}}{2^{m-1}} + \frac{\pi}{2^{2(m-1)}}\right)} - \sqrt{\cos\frac{\pi\tilde{r}}{2^{m-1}}}\right| \quad (30)$$

In summary, in Section 4.3, we have completely analyzed the success probability and calculation accuracy of QUIP, which are the key performance of QSNN. Thus far, the original minimum success probability $p_{\tilde{r}}$ for finding the local maximum potential is $\frac{4}{\pi^2}$. The maximum error $\varepsilon$ has a mapping relationship with the number of qubits $m$ and the measurement $\tilde{r}$ of the control register.

## 4.4 Improvement of the Minimum Success Probability

In the Section 4.3, we give the lower bound of the success probability and the maximum error, as shown in equation (28, 29) respectively. However, the success probability is too low (i.e., only $\frac{4}{\pi^2}$) to be acceptable. In this section, we want to further improve the minimum success probability, even if the calculation accuracy gets a slight deterioration. The accuracy loss can be compensated through increasing the number of qubits in control register.

Specifically, we use two approaches to improve the minimum success probability together. First, we now introduce a slack condition, we consider $\tilde{r}+1$, $\tilde{r}$, and $\tilde{r}-1$ as the correct result not just $\tilde{r}$. When $m$ is relatively larger, $r$ becomes larger accordingly, and this slack condition may cause a slight deterioration of the calculation accuracy. The probability $p$ to get $\tilde{r}+1$, $\tilde{r}$, and $\tilde{r}-1$ in a measurement of the control register is described as equation (31).

$$p = p_{\tilde{r}+1} + p_{\tilde{r}} + p_{\tilde{r}-1} \approx \frac{\sin^2(2^m\pi(\Delta r + 2^{-m}))}{2^{2m}(\pi(\Delta r + 2^{-m}))^2} + \frac{\sin^2(2^m\pi\Delta r)}{2^{2m}(\pi\Delta r)^2} + \frac{\sin^2(2^m\pi(\Delta r - 2^{-m}))}{2^{2m}(\pi(\Delta r - 2^{-m}))^2} \quad (31)$$

Similar to equation (27), we can obtain the minimum value of $p$, when $\Delta r \to 2^{-(m+1)}$ or $\Delta r \to -2^{-(m+1)}$.

$$\min p = \frac{76}{9\pi^2} \quad (32)$$

Additionally, repeatedly running the QUIP and measuring the control register $q$ times can help further improve the final success probability $p_q$ which is described as equation (33):

$$p_q = \sum_{k=\left[\frac{1}{2}q\right]+1}^{q} C_q^k (p)^k (1-p)^{q-k} \quad (33)$$

where $C_q^k = \frac{q!}{k!(q-k)!}$, $p_q$ denotes the success probability of obtaining the correct result from at least $q/2$ tests in repeated $q$ times. Obviously, if $p \geq 0.5$, $p_q$ will must be greater than $p$. This approach will cause $p_q$ close to 1 using a few copies of QUIP and the measurements, which especially is suitable for the case that the initial probability of a certain random event is greater than 0.5.

## 4.5 The Complexity of QSNN

For classic computers, the complexity for calculating an inner product of two vectors is $O(J)$, where $J$ is the dimension of each vector. Given the number of vector pairs as a constant $s$ (i.e., is related to the number of stimuli, layers and epochs of SNN), in the neural network model, the main complexity of SNN is $O(sJ)$. As for QSNN, we want to obtain the gate complexity of QUIP which is used to calculate the vector inner product. According to a proof on complexity of quantum phase estimation (QPE) algorithm proposed by [33], the complexity of QPE is polynomial to the complexity of its embedded black box operator $Q$. In our quantum circuits, the $Q$ is $U_G$. Since QPE and QAE share the same architecture, we can use the analysis of QPE's complexity for obtaining complexity of QAE. Therefore, we only need to further analyze the complexity of $U_G$. As for the four sub-operators of a $U_G$, the complexity of $O$ operator is $O(1)$; the complexity of $U_0$ operator (i.e., the complexity of a multi-controlled operator) is $O(\lceil \log_2 J \rceil^2)$ through adding the corresponding number of extra ancillary qubits [29]; the $U_s$ operator includes $\lceil \log_2 J \rceil$ controlled-swap gates, $U_w$ and $U_t$. $U_w$ and $U_t$ are essentially two operators used to prepare the initial quantum state $|w\rangle$ and $|t\rangle$, respectively. According to the previous studies, a polynomial complexity between the problem of preparing an initial quantum state and the number of qubits [16]. Therefore, in total, the complexity of a single test of QUIP is $O(\text{poly}\lceil \log_2 J \rceil)$. Considering the $s$, the complete complexity is $O(s \times \text{poly}\lceil \log_2 J \rceil)$. Consequently, the computational complexity of QSNN is log-polynomial, and is much lower than the linear computational complexity of the classic SNN.

## 5. Numerical Simulation and Application

In this section, we first verify the correctness of the equations (25,26,33), which are the key steps of QSNN, through a group of numerical simulations; Then, to verify the feasibility of QSNN, we do an image classification task using a real-world dataset.

## 5.1 Numerical Simulation

The schemes for the numerical simulations on the success probability and calculation accuracy of QSNN are as follows:
(1) We randomly generate 1000 pairs of $|w\rangle, |t\rangle$, and ensure their inner product result distribute in $[0,1]$, thus the numerical simulations can completely reflect the variations of $|w\rangle, |t\rangle$.
(2) We set $m$ qubits for the control register in QUIP, where $m = 4,6,8,10$. The simulation results of the success probability and the calculation accuracy are shown in Fig 7.
(3) To verify the approaches proposed in Section 4.5 to further improve the success probability, we set the repetitions of QUIP, $q = 1,3,5,7,9,11$. The simulation results of the success probability improvements are shown in Fig 8.

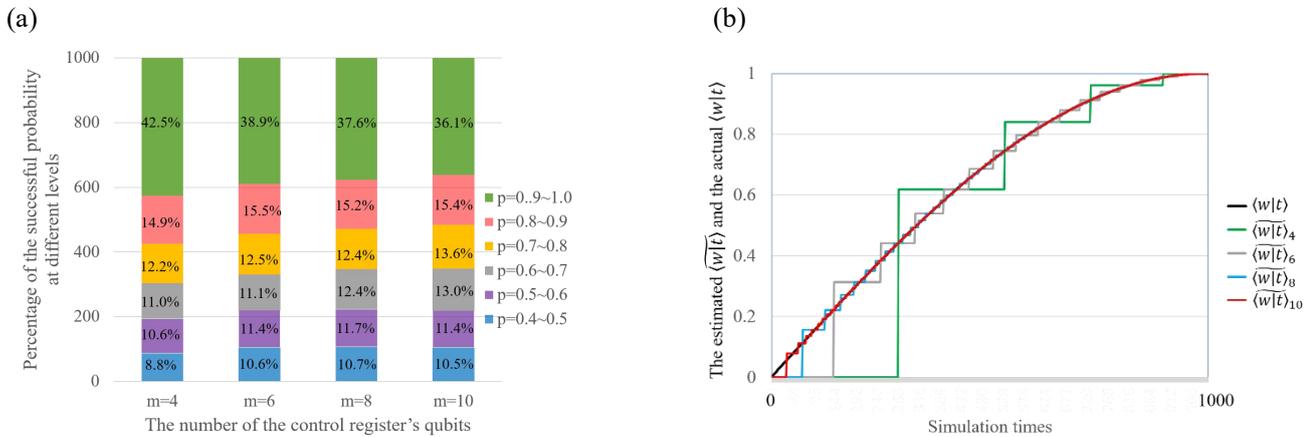

Fig 7. The simulations on 1000 vector pairs: (a) the success probability; (b) the calculation accuracy.

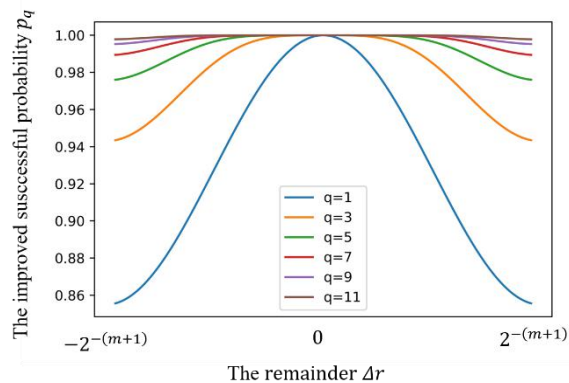

Fig 8. The simulation results of the success probability improvements

Fig 7(a) shows the success probability when the control register gives the correct result under different $m$. It can be concluded that, on one hand, the simulation results are in accord with the theoretical proof in equation (29), that is, the success probabilities of all $\langle w|t \rangle$ are greater than 0.4 $\left(\approx \frac{4}{\pi^2}\right)$. On the other hand, the number of control register qubits does not affect the success probability of QUIP. Fig 7(b) then shows the significant improvements of the calculation accuracy with the growth of $m$. From Fig 7(b), the control register can meet the usable calculation accuracy without a large number of qubits. Fig 8 shows the success probability that $\Delta r$ changes within the scope of the definition, i.e., $\Delta r \in \left[-2^{-(m+1)}, 2^{-(m+1)}\right)$. For the worst instance, when $q = 11$, $p = \frac{76}{9\pi^2}$, then the final success probability $p_q \approx 0.998$.

## 5.2  Image Classification by QSNN

In order to verify the feasibility of QSNN, we use a real-world dataset "MNIST" to do a image classification task. The MNIST originally consists of ten classes (i.e., the handwriting images from 0 to 9) and each sample is a grid of 28 × 28 pixels as shown in Fig 9. In this simulation, to reduce the implementation complexity, we only consider a binary classification, that is, the picture is identified as 0 or 1 through one spiking neuron. By adding more spiking neurons, QSNN can complete a multiple classification task, for example, we use the first spiking neuron to discriminate 0 and 1 through 9, and the second to discriminate 1 and 2 through 9, the rest is the same.

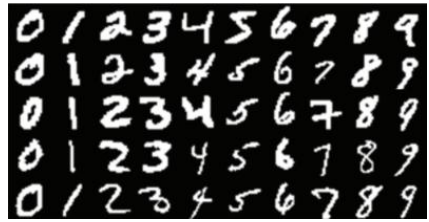

Fig 9. Images of the handwriting digits in MNIST.

The handwritten characters, a binary image matrix need to be encoded into a stimuli sequence as the inputs of QSNN. We briefly introduce the coding scheme. Specifically, a 28 × 28 image is input into a pooling layer and a 14 × 14 image is obtained. One row and one column are added to a 14 × 14 image to generate a 15 × 15 image; then, let each nine pixels of the 15 × 15 image be a stimulus as shown in Fig 10. In this way, the SNN has 25 synapses and each synapse has one temporal-spatial stimulus. The moment that the $i$-th synapse receives this stimulus is $t_i$. Then, $t_i = \sum_{k=0}^{8} 2^k \alpha_{i,k}$, where $\alpha_{i,k}$ represents the $k$-th pixel of the $i$-th synapse. Thus, we can transform the binary image (i.e., black and white) into different stimuli sequences. We use the weighted sum of the spikes on the synapses as the spiking neuron potential and classify the potentials over the threshold into '1', and others are classified into '0'.

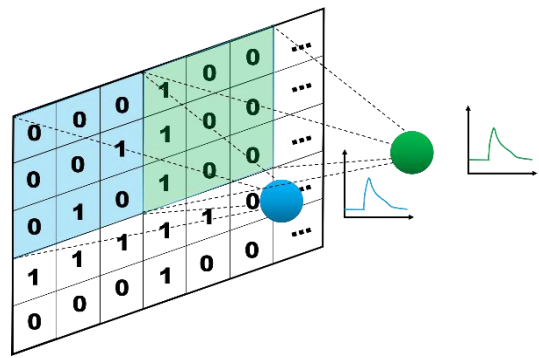

Fig 10. The encoding scheme for images of handwritten characters.

The encoded handwritten characters are input into the QSNN. Fig 11 displays the classification result for MNIST on a training set and a test set. Both the training and testing can achieve a high classification accuracy just in a few epochs of QSNN, and no obvious overfitting is found. Consequently, we can verify that QSNN is effective in real-world image classifications.

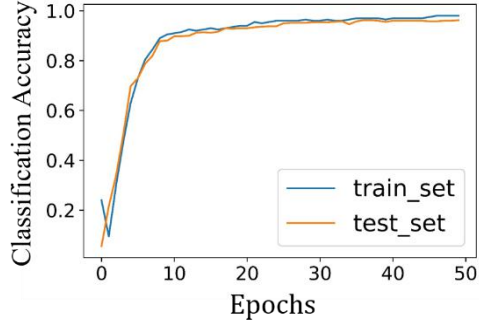

Fig 11. The classification accuracy in training and testing using QSNN.

## 6. Conclusion

In this paper, we proposed QSNN, in which we design and implement a quantum computing algorithm for the moment localizations of threshold crossings with highest computational complexity in the classic SNN. Specifically, we embedded the QUIP algorithm into the SNN, in which the control register with $m$ qubits is used to guarantee a high computation accuracy. In this way, $4^m$ runs and measurements of the traditional swap-test algorithm can be avoided. We gave a complete proof of the success probability and calculation accuracy of QSNN. Then, to further improve the minimum success probability, we proposed to use a slack condition and slight repetitions to improve the minimum success probability to nearly 100%. Through analyzation, it can be verified that, QSNN outperforms the classic SNN in computational complexity. The computational complexity of QSNN is log-polynomial to the data dimension, while the computational complexity of the classic SNN is linear to the data dimension. This suggests that our quantum algorithm can significantly accelerate SNNs, even for those with extremely high dimensions. To verify our mathematical proof of the minimum success probability and the calculation accuracy, we do numerical simulations by configuring different number $m$ of qubits in the control register. We found that $m$ not impact the minimum success probability, however, can significantly improve the calculation accuracy. This suggests that our QSNN can meet the computing requirements of arbitrary data accuracy. We also gradually increase the repetitions to see the improvements of the minimum success probability, and we found that when the repetitions $q = 11$, the minimum success probability reaches 0.998. Additionally, we use our QSNN to solve a real-world image classification problem, which verifies the feasibility of QSNN.

## Appendix A: Measurement Accuracy of Quantum State Probabilities in the Swap-test

An amplitude of a quantum state is usually used to store data information. The square of the amplitude is the measurement probability. In many cases, the probability partly contains the useful results. For instance, in this paper, the inner product is in the probability $P(|1\rangle)$ of the ancillary qubit in the swap-test. Any measurement of the ancillary qubit will cause this qubit collapse to 0 or 1 with the probability of $P(|0\rangle)$ or $P(|1\rangle)$, respectively. In this section, we give a mathematical proof that without the control register, the relationship between the accuracy of the $P(|1\rangle)$ measurement and $n_a$ repetitions of the swap-test run and measurement.

In order to obtain the probability of a qubit state, we must repeatedly execute the swap-test and count the measurement results. We regard each execution and measurement of the swap-test as a random event and an independent binomial distributed variable, respectively. Let the number of random events be $n_a$; let the set of variables be $X = \{x_1, x_2, \ldots, x_{n_a}\}$. Then the binomial distribution is represented as $S_{n_a} \sim B(n_a, P(|1\rangle))$. According to De Moivre–Laplace central limit theorem [34], if $n_a$ is large enough, for any real number $a, b$ $(a < b)$, the probability $P\left(a < \sum_{k=1}^{n_a} x_k \leq b\right)$ that $\sum_{k=1}^{n_a} x_k$ is in $(a,b]$ is described as equation (A1):

$$P\left(a < \sum_{k=1}^{n_a} x_k \leq b\right) \approx \Phi\left(\frac{b - n_a P(|1\rangle)}{\sqrt{n_a P(|1\rangle) P(|0\rangle)}}\right) - \Phi\left(\frac{a - n_a P(|1\rangle)}{\sqrt{n_a P(|1\rangle) P(|0\rangle)}}\right) \quad (A1)$$

where $\Phi(x)$ is standard Gaussian distributed function which is described as equation (A2).

$$\Phi(x) = \frac{1}{\sqrt{2\pi}} \int_{-\infty}^{x} e^{-\frac{y^2}{2}} dy \quad (A2)$$

In the swap-test, the interval (a, b] represents the lower and upper bound of the probability summation; $P\left(a < \sum_{k=1}^{n_a} x_k \leq b\right)$ represents the probability of $n_a P(|1\rangle) \in (a, b]$. Let $a = (1-\delta)n_a P(|1\rangle)$ and $b = (1+\delta)n_a P(|1\rangle)$. Thus, the calculation accuracy is referred to as δ. Then, equation (A1) can be rewritten as equation (A3).

$$P\left(a < \sum_{k=1}^{n_a} x_k \leq b\right) \approx 2\Phi\left(\delta\sqrt{n_a}\sqrt{\frac{P(|1\rangle)}{P(|0\rangle)}}\right) \quad (A3)$$

It can be found that δ and $(n_a)^2$ are inversely proportional, under the same probability condition from equation (A3).